\def\jpsi{\hbox{$J{\kern-0.24em}/{\kern-0.14em}\psi$}}
\def\ev#1#2{\hbox{#1e{\kern-0.10em}V{\kern-0.30em}/{\kern-0.14em}$#2$}}

\def\mevc{\ev{M}{c}}

\def\tbhead#1#2{Table\ #1.\  #2\vskip8pt}

\makeatletter
\def\eqalign#1{\null\,\vcenter{\openup\jot\m@th
  \ialign{\strut\hfil$\displaystyle{##}$&$\displaystyle{{}##}$\hfil
      \crcr#1\crcr}}\,}
\def\eqalignno#1{\displ@y \tabskip\@centering
  \halign to\displaywidth{\hfil$\@lign\displaystyle{##}$\tabskip\z@skip
    &$\@lign\displaystyle{{}##}$\hfil\tabskip\@centering
    &\llap{$\@lign##$}\tabskip\z@skip\crcr
    #1\crcr}}
\def\leqalignno#1{\displ@y \tabskip\@centering
  \halign to\displaywidth{\hfil$\@lign\displaystyle{##}$\tabskip\z@skip
    &$\@lign\displaystyle{{}##}$\hfil\tabskip\@centering
    &\kern-\displaywidth\rlap{$\@lign##$}\tabskip\displaywidth\crcr
    #1\crcr}}

\makeatother
\def\bqt#1#2\eqt{\begin{equation}\label{#1}%
        {#2}\end{equation}\noindent}
\def\bln#1#2\eln{\begin{equation}\label{#1}%
            \eqalign{#2}\end{equation}\noindent}
\def\brlist{\begin{thebibliography}{99}}
\def\brf{\bibitem}
\def\erlist{\end{thebibliography}}
\def\bra{\langle}
\def\ket{\rangle}
\documentclass{ws-p8-50x6-00}

\begin{document}

\title{                Measurement of\\
 Inclusive \lowercase{$f_1(1285)$} and \lowercase{$f_1(1420)$} Production \\
            in $Z$ Decays with the DELPHI Detector}

\author{Ph. Gavillet}

\address{EP Division, CERN, CH-1211 Gen\`eve 23, Switzerland\\
E-mail: Philippe.Gavillet@cern.ch\\[2mm]
               \mbox{\rm for}\\
               {\rm DELPHI Collaboration}}


\maketitle

\abstracts{
Inclusive production of two $(K\bar K\pi)^0$ states in the mass region
1.22--1.56 GeV in $Z$ decay at LEP I
has been observed by the DELPHI Collaboration.
The measured masses and widths are $1274\pm4$ and $29\pm12$ MeV for
the first peak and $1426\pm4$ and $51\pm14$ MeV for the second.
A partial-wave analysis has been performed on the $(K\bar K\pi)^0$
spectrum in this mass range; the first peak is consistent with
the quantum numbers $I^G(J^{PC})=0^+(0^{-+}/1^{++})$ 
and the second with $I^G(J^{PC})=0^+(1^{++})$.
These measurements, as well as their total hadronic production 
rates per hadronic $Z$ decay, are consistent with the mesons of the type
$n\bar n$, where $n=\{u,d\}$.  They are very likely to be
the $f_1(1285)$ and the $f_1(1420)$, respectively.
}
\section{Introduction}
The inclusive production of mesons has been a subject of long-standing
study at LEP\cite{delph3,uvr0} as it provides insight 
into the nature of fragmentation of quarks and gluons to hadrons.
For the first time, we present in this paper
a study of the inclusive production of two $J^{PC}=1^{++}$ mesons,
the $f_1(1285)$ and the $f_1(1420)$ (i.e. $^3P_1$).

There are at least four nonstrange isoscalar mesons,~\kern-4pt\cite{PDG}
$I^G(J^{PC})=0^+(1^{++})$ and $I^G(J^{PC})=0^+(0^{-+})$,
known in the mass region between 1.2 and 1.5 GeV, 
which couple strongly to the decay channel $(K\bar K\pi)^0$.  
They are $f_1(1285)$, $\eta(1295)$, $f_1(1420)$ and $\eta(1440)$, which
are mostly $n\bar n$ states, where $n=\{u,d\}$.
There exist possibly two additional states, $I^G(J^{PC})=0^-?(1^{+-})$
$h_1(1380)$ and $I^G(J^{PC})=0^+(1^{++})$ $f_1(1510)$, 
which may harbor a large $s\bar s$
content.~\kern-4pt\cite{PDG} 
Given this complexity in the $(K\bar K\pi)^0$ systems, it is
important that one seek answers as to which resonances among these 
are readily excited in inclusive hadron $Z$ decays.

The DELPHI data for this study is based on
the neutral $K\bar K\pi$ channel in the reaction
$Z\to (K_{_S}K^{\pm}\pi^{\mp})+X^0$.
\section{Experimental Procedure}
The analysis presented here is based 
on a data sample of about 3.3 million hadronic 
$Z$ decays collected from 1992 to 1995 
with the DELPHI detector.~\kern-4pt\cite{delph1,delph2} 

  The charged particle tracks have been measured 
in the 1.2-T magnetic field by a
set of tracking detectors. The average momentum resolution for charged
particles in hadronic final states, $\Delta p/p$, 
is usually between 0.001 and 0.01.

  A charged particle has been accepted 
in this analysis---if its momentum $p$ is
greater than 100 \mevc; its momentum error $\Delta p$ 
is less than $p$; and its impact
parameter with respect to the nominal crossing point is within 4 cm in the
transverse ($xy$) plane and 4 cm/$\sin{\theta}$ 
along the beam direction ($z$-axis),
$\theta$ being the polar angle of the track.

  Hadronic events are then selected by requiring at least 5 charged 
particles, with at least 3-GeV energy in each hemisphere of 
the event---defined with respect to the beam direction---and 
total energy at least 12\% of the center-of-mass energy. 

  After the event selection, in order to ensure a better signal-to-background
ratio for the resonances in the $K_{_S} K^\pm\pi^\mp$ 
invariant mass spectra, tighter requirements have been imposed 
on the track impact parameters, i.e. they have to be 
within 0.2 cm in the transverse plane 
and 0.4 cm/$\sin{\theta}$ along the beam direction.

  $K^\pm$ identification has been provided by the RICH detectors for
particles with momenta above 700 MeV/c, while the ionization loss measured in
the TPC has been used for momenta above 100 \mevc. 
Its efficiency has been estimated by comparing the experimental data
with simulated events generated with JETSET\cite{pyth4} tuned with
the DELPHI parameters\cite{delph5} and passed through the detector simulation
program DELSIM.~\kern-4pt\cite{delph6}

  The $K_{_S}$ candidates are detected by their decay in flight
into $\pi^+\pi^-$.  Our selection process consists of 
taking the $V^0$'s passing certain criteria\cite{delph7} 
for quality of the reconstruction plus a mass cut given by
0.45 $< M(K_{_S}) <$ 0.55 GeV.

   After all the above cuts, only events 
with at least one $K_{_S}K^+\pi^-$ or $K_{_S}K^-\pi^+$ combination 
have been kept in the present analysis, corresponding to
a sample of 705 688 events.
\section{$K_{_S}K^{\pm}\pi^{\mp}$ Mass Spectra}
\begin{figure}[ht]\vskip-12mm
\begin{center}\mbox{
\epsfig{file=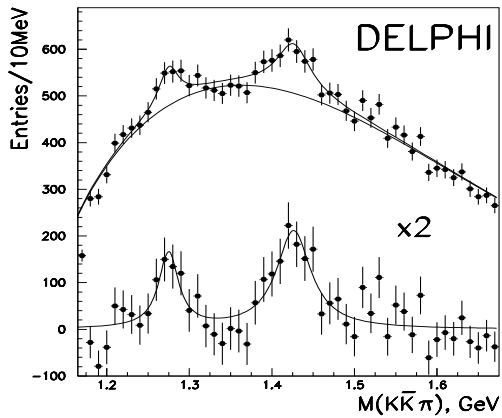,height=7.0cm,width=8.0cm}
}\end{center}
\vskip-4mm
\caption{$M(K_{_S}\,K^\pm\,\pi^\mp)$ distributions from the $Z$ decays
with the DELPHI detector at LEP I---with a mass cut
$M(K_{_S}\,K^\pm)<1.04$ GeV.  The two solid curves in the upper
part of the histogram describe Breit-Wigner fits over a smooth
background (see text).  The lower histogram and the solid
curve give the same fits with the background subtracted and amplified
by a factor of two.}
\label{fg02}
\end{figure}

The key to a successful study of the $f_1(1285)$ 
and $f_1(1420)$---given the enormous background in the 
$K_{_S}\,K^\pm\,\pi^\mp$ mass spectrum in this mass region---is 
to make a mass cut $M(K_{_S}\,K^\pm)\leq1.04$ GeV,
as shown in Fig.\,\ref{fg02}. Two clear peaks are seen in this
mass region.  There are two reasons for this: (1) the decay mode
$a_0(980)^\pm\pi^\mp$ is selected by the mass cut, while the general 
background for the $K\bar K\pi$ system is reduced by a factor of $\simeq7$
at 1.42 GeV or more at higher masses; (2) the interference effect of the
two $K^*(892)$ bands on the Dalitz plot at $M(K\bar K\pi)\sim1.4$ GeV
is enhanced, if the $G$-parity is positive.~\kern-4pt\cite{ch1}

   In order to measure the resonance parameters for these two
states, we have first generated a Monte Carlo sample, 
deleting---from the existing MC package---all mesons
with a major decay mode into $(K\bar K\pi)^0$ in the mass region 
1.25 to 1.45 GeV, i.e. $f_1(1285)$, $h_1(1380)$ and $f_1(1420)$,
which is then passed through the standard detector
simulation program. The smooth curve shown in Fig.\,\ref{fg02} has 
been obtained by fitting the mass spectrum of the aforementioned
MC sample between 1.15 to 1.65 GeV with a background function
\bln{s3a}
   f_b(M)=(M-M_0)^{\textstyle\alpha_1}\,
                \exp(\alpha_2\,M+\alpha_3\,M^2)
\eln  
where $M$ and $M_0$ are the effective masses of the $(K\bar K\pi)^0$ 
system and its threshold, respectively, and $\alpha_i$ are the 
experimental parameters.
We have fitted the $(K\bar K\pi)^0$ spectrum 
adding two $S$-wave Breit-Wigner forms 
to the background $f_b(M)$, given by
\bln{s3b}
   f_r(M)=\Gamma_r^2\Big/\left[(M-M_r)^2+(\Gamma_r/2)^2\right]
\eln
where $M_r$ and $\Gamma_r$ are the mass and the width to be determined
experimentally.
The results are shown in Fig.\,\ref{fg02} and also in Table I.
\vskip3mm
\noindent
\def\arraystretch{1.5}
\begin{center}
\begin{minipage}[]{100mm}
\tbhead{I}{Fitted parameters and numbers of events}
\begin{tabular}{|r@{$\,\pm\,$}l@{\hspace{12pt}}|
                     r@{$\,\pm\,$}l@{\hspace{12pt}}|
                                     r@{$\,\pm\,$}l@{\hspace{12pt}}|}
\hline\hline
\multicolumn{2}{|c|}{Mass (MeV)}&
\multicolumn{2}{c|}{Width (MeV)}&
\multicolumn{2}{c|}{Events}\\
\hline
\hspace{8pt}1274&4 & \hspace{16pt}29&12 
        & \hspace{8pt}345&$88\,({\rm stat})\pm69\,({\rm sys})$\\
\hspace{8pt}1426&4 & \hspace{16pt}51&14 
        & \hspace{8pt}790&$119\,({\rm stat})\pm110\,({\rm sys})$\\
\hline\hline
\end{tabular}
\end{minipage}
\end{center}
\def\arraystretch{1.0}
\vskip3mm
\noindent

    The main sources of systematic errors come from the various cuts and
selection criteria applied for the $V^0$ reconstruction plus the charged $K$ 
identification (7\%)---on the one hand---and 
the conditions of the mass-fit procedure---on the other (15\%). 
The systematic errors have been added quadratically 
and are shown in Table 1.  
\section{Partial-wave Analysis}
There exists a long list of 3-body partial-wave analyses; the reader
may consult PDG\cite{PDG} for earlier references, for example, on
$a_1(1260)$, $a_2(1320)$, $K_1(1270/1400)$ or $K_2(1770)$.  For the first
time, we apply the same technique to a study of the $(K\bar K\pi)^0$
system from the inclusive decay of the $Z$ at LEP.

   We have chosen to
employ the so-called Dalitz plot analysis, integrating over the three
Euler angles.~\kern-4pt\cite{ch0}  
The actual fitting of the data is done by using
the maximum-likelihood method, in which the normalization integrals
are evaluated with the accepted Monte Carlo events,~\kern-4pt\cite{e852a}
thus taking into account the finite acceptance of the detector
and the event selection.

The background under the two $f_1$'s is very large, some $\sim80\%$.
It is assumed that this represents essentially different processes
with, for example, different overall multiplicities---so that the
background does not interfere with the signals.  We assume further
that the background itself is a non-interfering superposition of
a flat distribution (on the Dalitz plot) and the partial waves
$I^G(J^{PC})=0^+(1^{++})\,a_0(980)\pi$,
$0^+(1^{++})\,(K^*(892)\bar K+c.c.)$ and $0^-(1^{+-})\,(K^*(892)\bar K+c.c.)$.
We have verified that these amplitudes give a good description of
the three background regions for $M(K\bar K\pi)$ in $1.22\to1.26$,
$1.30\to1.38$ and $1.48\to1.56$ GeV, respectively.

The signal regions, for $M(K\bar K\pi)$ in $1.26\to1.30$ and
$1.38\to1.48$ GeV, have been fitted with a non-interfering
superposition of the partial waves $I^G(J^{PC})=0^+(1^{++})$, 
$0^+(1^{+-})$ and $0^-(0^{-+})$,
where the decay channels $a_0(980)\pi$ and $K^*(892)\bar K+c.c.$ are
allowed to interfere within a given $J^{PC}$.  All other possible 
partial waves have been found to be negligible in the signal regions.
The fit results can be summarized as follows: 
(1) the maximum likelihood is found to be
the same for $I^G(J^{PC})=0^+(1^{++})\,a_0(980)\pi$ and for
$0^-(0^{-+})\,a_0(980)\pi$, i.e. the 1.28- GeV region is equally likely
to be the $f_1(1285)$ or the $\eta(1295)$; (2) in the 1.4-GeV region,
the maximum likelihood is better 
(by about 14 for $\Delta\ln{\cal L}$) for $I^G(J^{PC})=0^+(1^{++})$
$f_1(1420)$ than $I^G(J^{PC})=0^+(0^{-+})$ $\eta(1440)$; the
$I^G(J^{PC})=0^+(1^{+-})$ $h_1(1380)$ is excluded in this analysis
(by about 23 for $\Delta\ln{\cal L}$). 

   It should be emphasized that both the
mass-dependent (per 20 MeV) and the mass-independent global fits
give compatible results.  
\section{Discussion and Conclusions}
\begin{figure}[ht]\vskip2mm
\begin{center}\mbox{
\epsfig{file=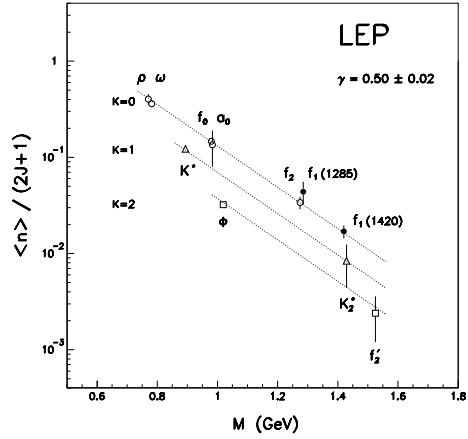,width=6.4cm}
}\end{center}
\vskip 3mm
\caption{Total production rate per spin state and isospin for scalar,
vector and tensor mesons as a function of the mass (open symbols).
The two solid circles correspond to the $f_1(1285)$ 
and the $f_1(1420)$.}
\label{fg04}
\vskip2mm
\end{figure}

We have measured the production rate $\bra n\ket$ 
per hadronic $Z$ decay for $f_1(1285)/\eta(1295)$ and $f_1(1420)$.
We assume for this study that {\em both have spin 1.}
The results are
\bln{hr0}
   \bra n\ket&=0.132\pm0.034\quad{\rm for}\quad f_1(1285)\cr
   \bra n\ket&=0.0512\pm0.0078\quad{\rm for}\quad f_1(1420)
\eln
taking a $K\bar K\pi$ branching ratio of $(9.0\pm0.4)$\% for the
$f_1(1285)$ and 100\% for the $f_1(1420)$.~\kern-4pt\cite{PDG}
The production rate per spin state [i.e. divided by $(2J+1)$] has been
studied;~\kern-4pt\cite{uvr0} in Fig.\,\ref{fg04} is given
all the available data for those mesons with a `triplet' $q\bar q$ 
structure, i.e. $S=1$ in the spectroscopic notation $^{2S+1}L_J$.
To this figure we have added our two mesons for comparison. 
It is seen that both $f_1(1285)$ and $f_1(1420)$ 
come very close to the line corresponding
to other mesons whose constituents are thought to be of the type $n\bar n$.
This is suggestive of two salient facts: (1)the first peak
at 1.28 GeV is very likely to be the $f_1(1285)$; (2) both
$f_1(1285)$ and $f_1(1420)$ have little $s\bar s$ content.

   We have studied the inclusive production of 
$f_1(1285)/\eta(1295)$ and $f_1(1420)$ in $Z$ decays at LEP I.
The measured masses and widths are $1274\pm4$ and $29\pm12$ MeV for
the first peak and $1426\pm4$ and $51\pm14$ MeV for the second one.
For the first time, a partial-wave analysis has been carried out
on the $(K\bar K\pi)^0$ system.  The results show
that the first peak is equally likely to be the $f_1(1285)$
or the $\eta(1295)$, while the second peak is consistent 
with the $f_1(1420)$.  However, the hadronic production rate of
these two states suggests that their quantum numbers are very probably
$I^G(J^{PC})=0^+(1^{++})$ and that their quark constituents are mainly
of the type $n\bar n$, where $n=\{u,d\}$.  
Finally, we conclude that the mesons
$\eta(1295)$, $\eta(1440)$ and $h_1(1380)$ are less likely to be produced
in the inclusive $Z$ decays compared to $f_1(1285)$ and $f_1(1420)$.
\vskip3mm
\brlist
\bibliographystyle{unsrt}
\brf{delph3} P. Abreu {\it et al.}, 
             DELPHI Collab., Phys. Lett. {\bf B449} (1999) 364.
\brf{uvr0} V. Uvarov, Phys.Lett. B511 (2001) 136.
\brf{PDG}  Review of Particle Physics,
                  Eur. Phys. J. C{\bf 15}, (2000) 1.
\brf{delph1}  P. Aarnio {\it et al.}, 
             DELPHI Collab., Nucl. Inst. Meth. {\bf A303} (1991) 233.
\brf{delph2}  P. Abreu {\it et al.}, 
             DELPHI Collab., Nucl. Inst. Meth. {\bf A378} (1996) 57.
\brf{pyth4}  T. Sj\"{o}strand, Comput. Phys. Comm. {\bf 82} (1994) 74.
\brf{delph5}  P. Abreu {\it et al.}, DELPHI Collab.,
                                         Z.Phys. C 73 (1996) 11.
\brf{delph6} P.Aarnio {\it et al.}, DELPHI Collab.,
                                 Nucl. Instr. Meth. A 378 (1996) 57.
\brf{delph7} P. Abreu {\it et al.}, DELPHI Collab.,
                                         Z.Phys. C 65 (1995) 587.
\brf{ch1} See S. U. Chung, 
 `Analysis of $K\bar K\pi$ systems (Version I),' BNL-QGS-98-901 (1998), 
         http://cern.ch/suchung/.
\brf{ch0} See S. U. Chung, 
               `Spin Formalisms,' CERN preprint 71-8 (1971).
\brf{e852a} S. U. Chung {\it et al.}, 
                    Phys. Rev. {\bf D60} (1999) 092001.
\erlist
\end{document}